\documentclass[a4paper,11pt]{article}
\usepackage{pos}
\usepackage{subcaption}
\usepackage{mathtools}
\usepackage[numbers]{natbib} 
\usepackage{blindtext}
\newcommand{\Htwo}[0]{\ensuremath{\mathrm{H}_2}}%

\title{Can cosmic rays explain the high ionisation rates in the Galactic centre?}
\ShortTitle{Can CR explain the high ionisation rate in the GC?}

\author*[a]{Sruthiranjani Ravikularaman}
\author[b]{Sarah Recchia}
\author[c]{Vo Hong Minh Phan}
\author[d]{Stefano Gabici}

\affiliation[a]{Astronomisches Institut (AIRUB), Ruhr-Universität Bochum, Germany}
\affiliation[b]{INAF-Osservatorio Astronomico di Brera, Via Bianchi 46, I-23807 Merate, Italy}
\affiliation[c]{Sorbonne Université, Observatoire de Paris, PSL Research University, LERMA, CNRS UMR 8112, 75005 Paris, France}
\affiliation[d]{Université Paris Cité, CNRS, Astroparticule et Cosmologie, F-75013 Paris, France}

\emailAdd{sruthi@astro.ruhr-uni-bochum.de}

\abstract{The Central Molecular Zone (CMZ), located in the centre of the Milky Way, is a roughly cylindrical structure of molecular gas extending up to  parsecs around the supermassive black hole Sagittarius A$^{*}$. The average $\Htwo$ ionisation rate in the CMZ is estimated to be $2\times 10^{-14}~\rm{s}^{-1}$, which is 2–3 orders of magnitude higher than anywhere else in the Galaxy. Due to the high gas density in this region, electromagnetic radiation is rapidly absorbed, leaving low-energy cosmic rays (CRs) as the only effective ionising agents. Hence, a high CR density has been invoked to explain such high ionisation rates. However, a corresponding excess in $\gamma$-rays, which would result from interactions of high-energy CRs, has not been observed. This suggests that the supposed excess exists only in the low-energy CR spectrum. To constrain this unknown low-energy component, we first derive the high-energy CR injection spectra using $\gamma$-ray and radio data, to which we add various low-energy components. We then propagate these injection spectra by numerically solving the CR transport equation using a Crank-Nicolson scheme. Testing multiple CR injection scenarios, we find that the energy required to sustain the observed ionisation rates is excessively high in every case. We conclude that CRs cannot be the exclusive ionising agents in the CMZ.}

\ConferenceLogo{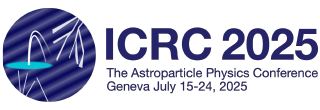}

\FullConference{39th International Cosmic Ray Conference (ICRC2025)\\
 15–24 July 2025\\
Geneva, Switzerland\\}

\begin{document}
\maketitle

\section{Introduction}

The central molecular zone (CMZ) is a roughly cylindrical volume surrounding the supermassive black hole (SMBH) Sagittarius A$^*$ (Sgr A$^*$), that has a radius of $\approx200~\mathrm{pc}$ and a height of $\approx 100$~pc and a mass of $M_\mathrm{CMZ}\sim6\times10^{7}~\mathrm{M}_\odot$ \cite{fer07, tsu99}. The $\Htwo$ ionisation rate in this region is estimated to be $\zeta_\mathrm{CMZ}=2\times10^{-14}~\mathrm{s}^{-1}$ \cite{oka19}. Although, both cosmic rays (CRs) and photons (UV, X-rays) can ionise $\Htwo$, it is believed that the former is primarily responsible for the ionisation rate in this region as ionising photons cannot penetrate large column densities of gas \citep{mck89}. Among CRs, those with particle energy in the sub-giga-electronvolt domain are believed to be most effective in ionising interstellar matter \citep[see e.g.][and references therein]{gab22}.

The ionisation of molecular hydrogen leads to the production of H$_3^+$ ions. By balancing the rates of formation and destruction of $\rm H_3^+$, we obtain an expression for the ionisation rate:
\begin{equation}
    \zeta_\mathrm{CR}^{\Htwo} = \frac{2k_\mathrm{e}x_\mathrm{e}}{f_{\Htwo}}\frac{N({\rm H}_3^+)}{L}
    \label{zL}
,\end{equation} where $x_\mathrm{e}$ is the electron fraction, $f_{\Htwo}=2n(\Htwo)/n_{\rm H}$ is the fraction of molecular hydrogen, and $N({\rm H}_3^+)$ is the H$_3^+$ column density along the line of sight of length $L$.
Other ionisation processes, such as double or dissociative ionisation, are negligible compared to these processes \citep{pad09} and will be ignored in the following. 

Following \cite{pad09}, the CR ionisation rate from protons and electrons can be expressed as a function of the CR proton and electron particle distribution functions ($f_\mathrm{p}$ and $f_\mathrm{e}$, respectively):
\begin{equation}
\begin{split}
    \zeta_\mathrm{p} &= \int_{E_\mathrm{min}}^{E_\mathrm{max}}v_\mathrm{p}f_\mathrm{p}(E_\mathrm{p})\sigma_\mathrm{p}^{ion}(E_\mathrm{p})(1+\phi_\mathrm{p}(E_\mathrm{p}))\,dE_\mathrm{p} \\
    &+ \int_{0}^{E_\mathrm{max}}v_\mathrm{p}f_\mathrm{p}(E_\mathrm{p})\sigma_\mathrm{p}^{e.c.}(E_\mathrm{p})\,dE_\mathrm{p} \\
    \zeta_\mathrm{e} &= \int_{E_\mathrm{min}}^{E_\mathrm{max}}v_\mathrm{e}f_\mathrm{e}(E_\mathrm{e})\sigma_\mathrm{e}^{ion}(E_\mathrm{e})(1+\phi_\mathrm{e}(E_\mathrm{e}))\,dE_\mathrm{e} 
\end{split}
\label{eq:zeta}
,\end{equation} 
where $v_\mathrm{i}$ is the particle velocity, 
and $\phi_\mathrm{i}$ the average number of secondary ionisations per primary ionisation, modelled as in \cite{kra15}.
Available parameterisations of the proton impact ionisation cross-section, $\sigma_\mathrm{p}^{ion}$, the electron capture cross-section, $\sigma_\mathrm{p}^{e.c.}$, and the electron impact ionisation cross-sections, $\sigma_\mathrm{e}^{ion}$ \cite{pad09, kra15, gab22}. 
The impact of CR nuclei heavier than protons should also be accounted for when computing the CR ionisation rate. The CR ionisation rate of CR nuclei (including protons) is $\zeta_\mathrm{n}=\eta\zeta_\mathrm{p}$ where the enhancement factor is $\eta\approx1.5$ \cite{pad09}.

In this paper, we present results from a detailed study on the plausibility of the CR ionisation model to explain the very large ionisation rates observed in the CMZ \citep{rav25}. 

\section{Cosmic-ray transport and spectra in the CMZ}

The equation describing the evolution in time, space, and momentum of the CR particle distribution function, $f(t,r,z,p)$, is \citep{ber90}:

\begin{equation}
\begin{split}
    \frac{\partial{f}}{\partial{t}} &= D_{rr}\frac{\partial^2{f}}{\partial{r^2}} + \left(\frac{D_{rr}}{r} + \frac{\partial{D_{rr}}}{\partial{r}}\right)\frac{\partial{f}}{\partial{r}} + D_{zz}\frac{\partial^2{f}}{\partial{z^2}} + \frac{\partial{D_{zz}}}{\partial{z}}\frac{\partial{f}}{\partial{z}} - v_\mathrm{w}\frac{\partial{f}}{\partial{z}}\\
    &+ \frac{1}{p^2}\frac{\partial}{\partial{p}}\left(p^2D_{pp}\frac{\partial{f}}{\partial{p}}\right) + \frac{p}{3}\frac{\partial{v_\mathrm{w}}}{\partial{z}}\frac{\partial{f}}{\partial{p}}  - \frac{1}{p^2}\frac{\partial}{\partial{p}}(\dot{p}p^2f) + Q
\end{split}
\label{eq:transport}
,\end{equation} 
where $D_{rr}$ and $D_{zz}$ are the CR spatial diffusion coefficients along $r$ and $z$, respectively, $v_\mathrm{w}$ is the Galactic wind velocity (assumed to be directed along the z axis), $D_{pp}$ the CR diffusion coefficient in momentum (particle re-acceleration), $\dot{p}$ the momentum loss rate, and $Q$ the particle injection rate. In the following, we consider the case of an isotropic and spatially uniform diffusion ($D(p) = D_{rr} = D_{zz}$) and a wind speed that is independent of $z$. This equation is solved using our custom particle transport equation solver that is based on a Crank-Nicolson scheme (see \cite{rav25} for a detailed explanation on the choice of parameters and the numerical solver). 

Following \cite{HESS2016}, the high-energy ($p>p_t$) CR injection is assumed to be central and continuous and the spectra are assumed to be power laws (or broken power laws) in momentum, giving:
\begin{equation}
\label{eq:injection1}
    Q_\mathrm{i}(r, z, p) = Q_\mathrm{p,i}(p)\frac{\delta(r)}{2\pi{r}}\delta(z)
,\end{equation} where 
\begin{equation}
\label{eq:injection2}
    Q_\mathrm{p,i}(p) = Q_\mathrm{0,i}\left(\frac{pc}{10~\mathrm{TeV}}\right)^{-\delta_\mathrm{i}}
,\end{equation} where the subscript $i$ can refer to either protons ($i = p$) or electrons ($i = e$).

By fitting the expected $\gamma$-ray and radio emissions \cite{HESS2006, HESS2016, gag17, law08} from the stationary spectra resulting from the transport of these injection spectra, we constrain the spectral indices and normalisations for the CR injection of particles with $p>p_{t}$ ($p_tc=0.78~\rm{GeV}$ for protons corresponding to a kinetic energy equal to the pion production threshold and $p_tc=0.45~\rm{GeV}$ for electrons). The CR proton spectrum normalisation and spectral index are:

\begin{eqnarray}
Q_\mathrm{p,p}(10~\mathrm{TeV}) &=& 1.1\times10^{29}~\mathrm{MeV}^{-1}~\mathrm{s}^{-1} \\
\delta_\mathrm{p} &=& 4.1
.\end{eqnarray}

In the case of the CR electron, we assume a broken power-law:

\begin{equation}
        Q_\mathrm{p,e}(p)= 
\begin{dcases}
    Q_*\left(\frac{p}{p_*}\right)^{-\delta_\mathrm{e,1}} ,& \text{if } p\leq p_*,\\
    Q_{*}\left(\frac{p}{p_*}\right)^{-\delta_\mathrm{e,2}} ,& \text{otherwise}
\end{dcases}
\label{eq:bknpl}
\end{equation} where $p_*c=1.5~\mathrm{GeV}$.
The normalisation is:

\begin{equation}
    Q_\mathrm{p,e}(10~{\rm TeV}) = 1.5\times10^{27}~\mathrm{MeV}^{-1}~\mathrm{s}^{-1}
,\end{equation}
and the spectral indices are:
\begin{eqnarray}
    \delta_\mathrm{e,1} &= 3.25\\
    \delta_\mathrm{e,2} &= 4.4 
\end{eqnarray}

To the contribution from the CRs accelerated in the GC, we add that from the background Galactic CRs. We assume the background CRs to have the same intensity as the ones measured in the local interstellar medium (the label LIS stands for local interstellar spectrum) by Voyager and AMS-02 \citep[see e.g.][and references therein]{gab22}. In the following, we explore different CR injection scenarios in the low-energy regime ($p<p_t$) which is responsible for the ionisation.

\section{Continuous injection scenarios}

\begin{figure*}[htbp]
    \begin{subfigure}{.5\linewidth}
        \includegraphics[width=\linewidth]{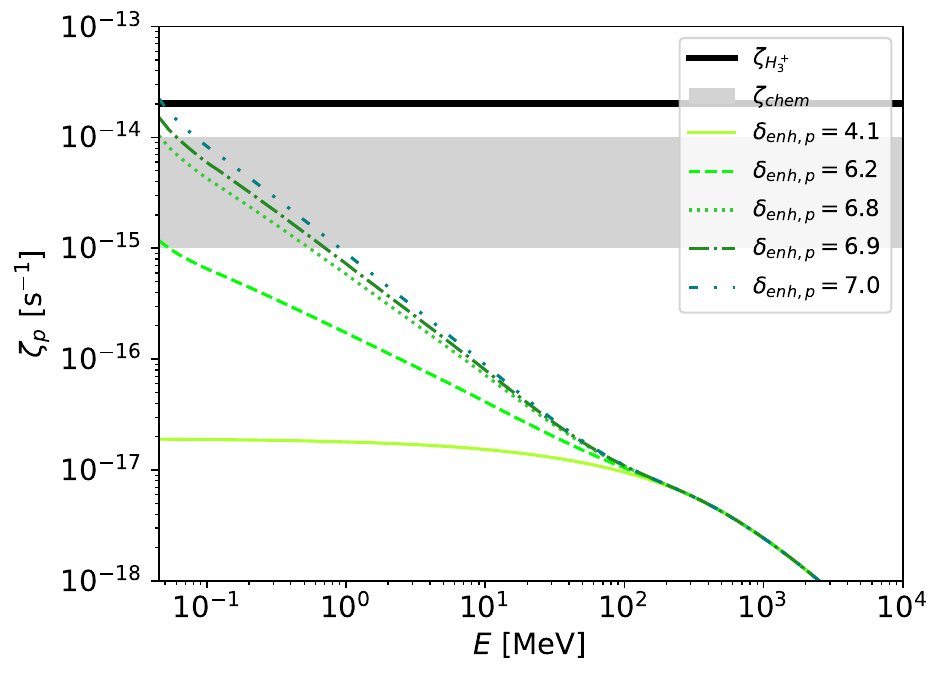}
    \end{subfigure}
    \hfill
    \begin{subfigure}{.5\linewidth}
        \includegraphics[width=\linewidth]{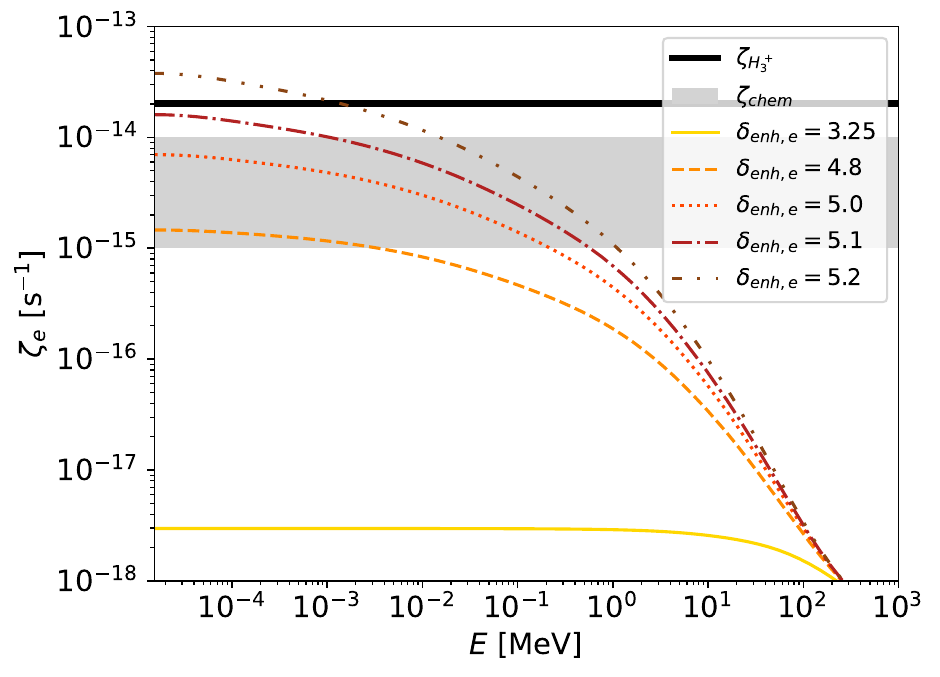}
    \end{subfigure}
      \caption[Average ionisation rates for different enhancements in the CMZ as a function of minimum particle energy]{Average ionisation rates in the CMZ as a function of minimum particle energy \textbf{Left:} for protons and \textbf{Right:} for electrons and compared to the estimated values of ionisation rates derived by different observational methods (horizontal black line and shaded region).}
    \label{fig:ion}
\end{figure*}

The $\gamma$-ray observations from the CMZ have hinted towards the existence of a continuous CR injection in the centre. Therefore, we first consider that these sources also supply the CRs responsible for the high ionisation rates. A simple extrapolation of the high-energy spectra obtained from fitting observations gives ionisation rates that are too low compared to the measurements in the CMZ. Hence, we expect an extra component on top of the continuous injection spectra suggested by $\gamma$-ray observations. We do so by adding a steeper power-law in the low energies (below 780 MeV for protons and 450 MeV for electrons, so as to not contradict $\gamma$-ray and radio observations). Fig.~\ref{fig:ion} shows the ionisation rate as a function of minimum ionising CR energy for different slopes of the enhancement. 

We see from Fig.~\ref{fig:ion} that in order to reproduce the ionisation rate measurements, one needs an extremely steep additional CR component (slopes of the order $\delta_{\rm enh,p} \sim 7$ and $\delta_{\rm enh,e} \sim 5$ for CR protons and electrons, respectively) extending down to very low particle energies ($E_{\rm min} \approx$~1 keV-1 MeV for both CR protons and electrons). These spectra correspond to a power of $6.8\times10^{40}~\mathrm{erg}\,\mathrm{s}^{-1}$ for protons and $2.4\times10^{40}~\mathrm{erg}\,\mathrm{s}^{-1}$ for electrons. Compared to the total CR power in the Galaxy, $\sim 6.0 - 7.4 \times10^{40}~\mathrm{erg}\,\mathrm{s}^{-1}$ and $\sim 1.0 - 1.7 \times 10^{39}~\mathrm{erg}\,\mathrm{s}^{-1}$ in CR protons and electrons, respectively \cite{str10}, the powers needed here are very large.

\begin{figure*}[htbp]
    \begin{subfigure}{.5\linewidth}
        \includegraphics[width=\linewidth]{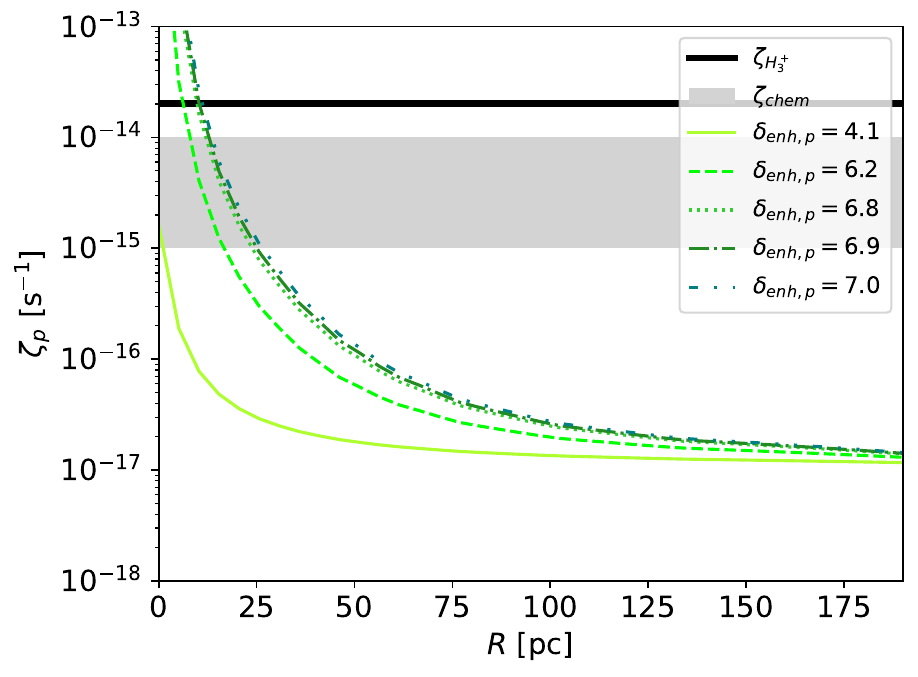}
    \end{subfigure}
    \hfill
    \begin{subfigure}{.5\linewidth}
        \includegraphics[width=\linewidth]{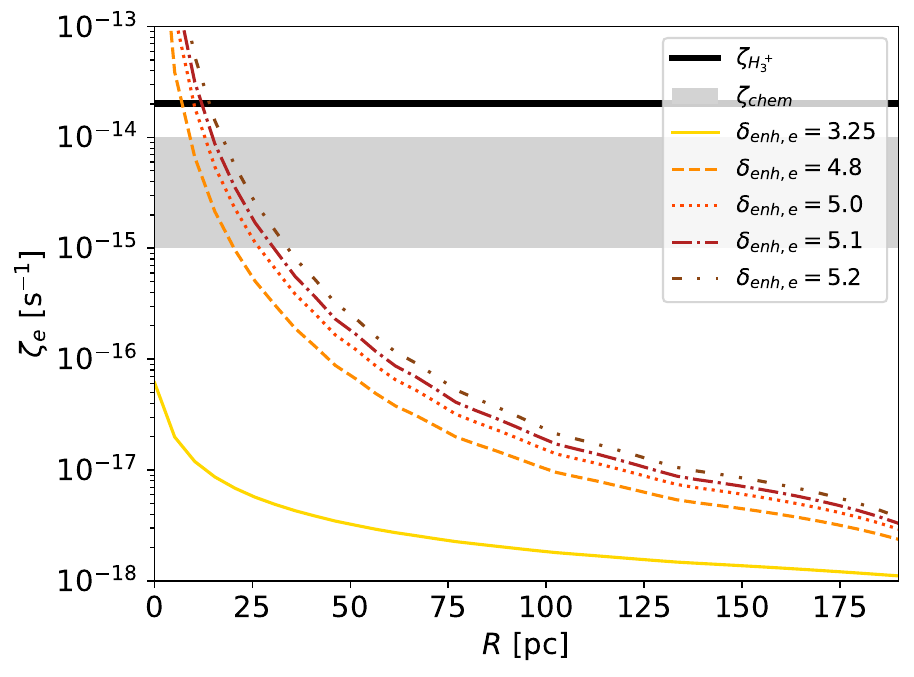}
    \end{subfigure}
      \caption[Average ionisation rates for different enhancements as a function of galactocentric distance]{Ionisation rates for different enhancements averaged over lines of sight at different projected distances $R$ from the GC in the CMZ as a function of minimum particle energy \textbf{Left:} for protons and \textbf{Right:} for electrons and compared to the estimated values of ionisation rates derived by different observational methods (horizontal black line and shaded region).}
    \label{fig:ionR}
\end{figure*}

Moreover, another property of the observed ionisation rates that should be reproduced is the lack of a spatial variation along the Galactocentric radius. Our model cannot reproduce this trend. In Fig.~\ref{fig:ionR}, we show the CR ionisation rate averaged along the lines of sights laying along the GP at different distances $R$ from the GC.

Next, we imagine the existence of a low-energy component that may not be related to the continuously injected spectra of CRs responsible for the $\gamma$-ray and radio observations. This low-energy distribution must have a cutoff at the transition energies used above. The spectral shape may range from a monoenergetic injection to a flat spectrum. 

In the case of a monoenergetic delta-injection, the possible choice for the energy at which we should increase the CR density is the energy with the highest probability of ionising interaction. We choose a delta-injection represented by a narrow Gaussian peaking at the energy where the proton and electron ionisation cross sections peak, respectively. Hence, we have a delta-injection for protons, peaking at $p\simeq63~\mathrm{keV}$ with a width $\sigma=0.1~\mathrm{MeV}$. Similarly, we take a delta-injection for electrons, peaking at $p\simeq69~\mathrm{eV}$ with a width $\sigma=0.5~\mathrm{keV}$. The normalisation is enhanced until the needed ionisation rate is reached. The CR powers corresponding to delta-injections that reach the observed ionisation rates are $3.6\times10^{41}~\mathrm{erg}\,\mathrm{s}^{-1}$ for protons and $2.3\times10^{40}~\mathrm{erg}\,\mathrm{s}^{-1}$. A delta-injection is even more energetically costly than the steep enhancements added in the previous section. Therefore, we may rule out a monoenergetic injection scenario to explain the high ionisation rates.

We also test flat spectra that have a steep cut-off before the transition energies. In this case, no matter the spectral index, the power needed to supply CR protons is always more than $10^{40}~\mathrm{erg}~\mathrm{s}^{-1}$. In the case of electrons, the total power is acceptable for some spectral slopes, but the flux of MeV $\gamma$-rays from the CMZ would be higher than the flux from point sources that have been observed by \textit{SPI} and \textit{COMPTEL} and therefore must have been observed, which is not the case.

Until now, we have only considered central injections. In the optimistic case of CR sources injecting particles everywhere in the CMZ, we are able to reproduce the spatial variation of the ionisation rate. However, the energetic cost remains high: $6.8\times 10^{40}~\mathrm{erg}~\mathrm{s}^{-1}$ for protons and $3.2\times 10^{40}~\mathrm{s}^{-1}$. The power we require is many factors larger than the total mechanical power injected by supernovae in this region \cite{cro11}. Moreover, in this injection scenario, this enormous CR power may not be associated with the SMBH, making these values even more inexplicable. 

\section{Impulsive injection scenarios}

In the previous section, an underlying assumption is that these high ionisation rates do not vary over time. Hence, we only explored the scenarios where particles are injected continuously. In this section, we derive constraints on a hypothetical powerful outburst. In particular, we compute the total energy that needed to be injected and how recently this injection should have taken place in order to observe ionisation rates $\gtrsim10^{-14}~\mathrm{s}^{-1}$. As the logic followed in the previous section, this impulsive injection should only introduce low-energy particles in the region. 

We choose the enhancement slopes from the continuous section that could reach the observed ionisation rates. This corresponds to $\delta_{\rm p}=7.0$ for CR protons and $\delta_{\rm e}=5.2$ for CR electrons. We assume that the injection event took place at a time $t_0$.

\begin{figure*}[htbp]
    \begin{subfigure}{.5\linewidth}
        \includegraphics[width=\linewidth]{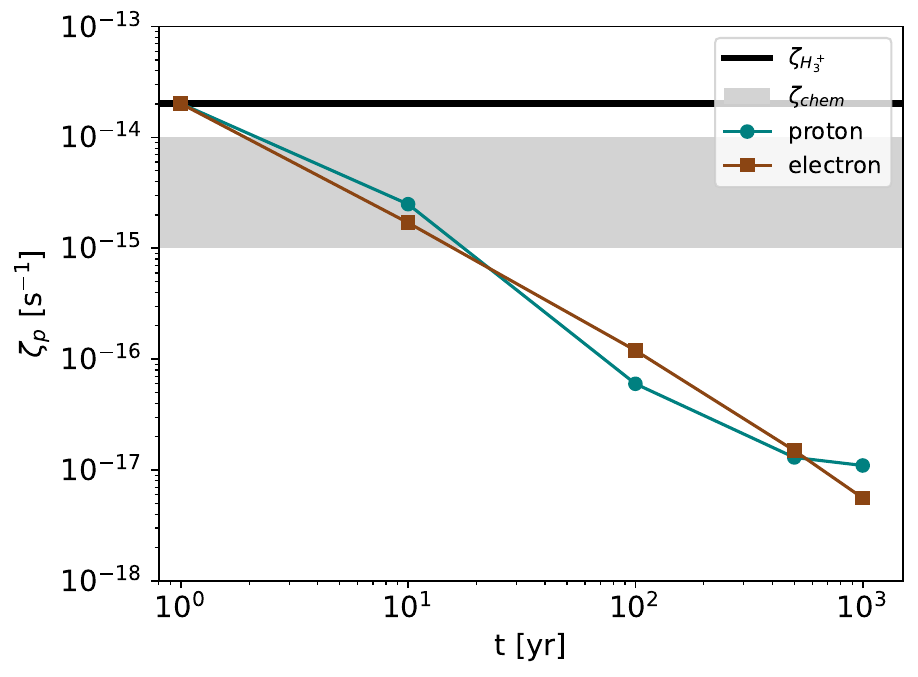}
    \end{subfigure}
    \hfill
    \begin{subfigure}{.5\linewidth}
        \includegraphics[width=\linewidth]{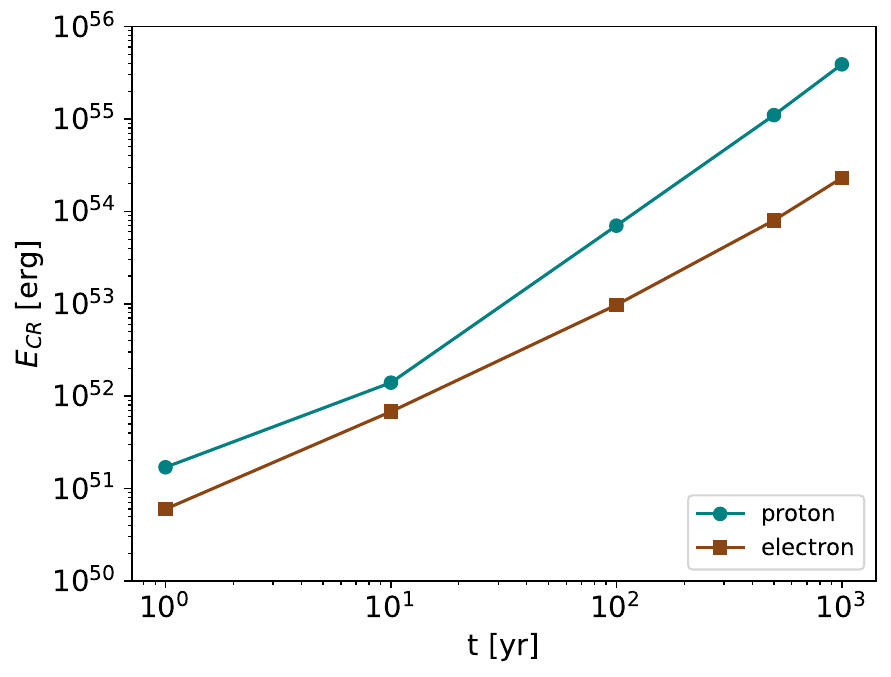}
    \end{subfigure}
      \caption[Ionisation rates from an impulsive injection]{\textbf{Left:} Evolution of ionisation rates from an impulsive injection at $t=0$ in time and \textbf{Right:} CR energy required in an outburst $t$ years ago for the ionisation rate to be above $10^{-14}~\rm{s}^{-1}$.}
    \label{fig:zetaimp}
\end{figure*}

The left panel of Fig.~\ref{fig:zetaimp} shows that a high ionisation rate resulting from an impulsive injection drops by an order of magnitude within the first ten years. However, the earliest measurements of the high ionisation rate date back at least $\sim~$20 years. The right panel of Fig.~\ref{fig:zetaimp} shows the amount of energy that needed to be released in CR at an outburst that happened t years ago, for the ionisation rate to be at least $2\times 10^{-14}~\rm{s}^{-1}$ at present. If we require the ionisation rate to be at least $2\times 10^{-14}~\rm{s}^{-1}$ during the last 20 years, an outburst releasing more than $10^{52}~\rm{ergs}~\rm{s}^{-1}$ should have occurred 20 years ago. This is highly unlikely as the most recent outburst at the GC is supposed to have been a few 100 years ago \cite{pon13}. Assuming this outburst lasted $\sim 10$ years, the total energy injected was of the order $\sim 10^{47-48}~\mathrm{ergs}$. A more recent and powerful event would have been observed, which has not been the case. Any earlier injection increases the associated energy proportionately. Therefore, an impulsive injection scenario is also not capable of explaining the high ionisation rates.

\section{Conclusion}

The ionisation rate in the CMZ has been estimated to be $2\times 10^{-14}~\mathrm{s}^{-1}$, which is a few orders of magnitude higher than the ionisation rates everywhere else in the Galaxy. Since photons would be quickly absorbed by the high column densities of gas in this region, CRs are assumed to cause this excess in the ionisation rate. Considering various CR injection scenarios in the GC region, we have come to the conclusion that it is extremely unlikely that CRs in the CMZ are the agents responsible for the large ionisation rates derived from a number of observations \cite{rav25}.  

Recently, \cite{obo24} used the latest dust extinction maps and revised C$_2$ data to obtain the density distribution in parsec scales towards diffuse clouds. Reevaluated ionisation rates along these sightlines have decreased by an order of magnitude compared to previous estimations. Therefore, we can not ignore the possibility that the ionisation rate in the GC may need to be reevaluated. However, the gas column length $L$ assumed in \cite{oka19} is 100 pc. $L$ can at most be the entire diameter of the CMZ which will only be a few factors higher, making the ionisation drop by a few factors. However, the CR ionisation scenario becomes plausible only if the ionisation rate is $10^{-15}~\rm{s}^{-1}$ or less. Hence, the issue of the high ionisation rates in the GC remains. Moreover, \cite{rav25} showed that for any value of $L$, it is impossible to find such high ionisation rates caused by CRs unless the star used for H$_3^+$ absorption is directly behind the SMBH, in which case it might not even be visible. 

In conclusion, CRs can not explain the high ionisation rates observed in the CMZ. The most obvious alternative candidate for ionisation is a radiation field made of UV and/or X-ray photons. If photoionisation proves to be ineffecient, exotic phenomena, such as ionisation by dark matter, may need to be explored.

\bibliography{refs}

\providecommand{\href}[2]{#2}\begingroup\raggedright\begin{thebibliography}{10}

\bibitem{fer07}
K.~{Ferri{\`e}re}, W.~{Gillard} and P.~{Jean}, \emph{{Spatial distribution of interstellar gas in the innermost 3 kpc of our galaxy}}, \href{https://doi.org/10.1051/0004-6361:20066992}{\emph{A\&A} {\bfseries 467} (2007) 611} [\href{https://arxiv.org/abs/astro-ph/0702532}{{\ttfamily astro-ph/0702532}}].

\bibitem{tsu99}
M.~{Tsuboi}, T.~{Handa} and N.~{Ukita}, \emph{{Dense Molecular Clouds in the Galactic Center Region. I. Observations and Data}}, \href{https://doi.org/10.1086/313165}{\emph{ApJS} {\bfseries 120} (1999) 1}.

\bibitem{oka19}
T.~{Oka}, T.R.~{Geballe}, M.~{Goto}, T.~{Usuda}, {Benjamin}, J.~{McCall} et~al., \emph{{The Central 300 pc of the Galaxy Probed by Infrared Spectra of H$_3^+$ and CO. I. Predominance of Warm and Diffuse Gas and High H$_{2}$ Ionization Rate}}, \href{https://doi.org/10.3847/1538-4357/ab3647}{\emph{ApJ} {\bfseries 883} (2019) 54} [\href{https://arxiv.org/abs/1910.04762}{{\ttfamily 1910.04762}}].

\bibitem{mck89}
C.F.~{McKee}, \emph{{Photoionization-regulated Star Formation and the Structure of Molecular Clouds}}, \href{https://doi.org/10.1086/167950}{\emph{ApJ} {\bfseries 345} (1989) 782}.

\bibitem{gab22}
S.~{Gabici}, \emph{{Low energy cosmic rays}}, {\emph{arXiv e-prints} (2022) arXiv:2203.14620} [\href{https://arxiv.org/abs/2203.14620}{{\ttfamily 2203.14620}}].

\bibitem{pad09}
M.~{Padovani}, D.~{Galli} and A.E.~{Glassgold}, \emph{{Cosmic-ray ionization of molecular clouds}}, \href{https://doi.org/10.1051/0004-6361/200911794}{\emph{A\&A} {\bfseries 501} (2009) 619} [\href{https://arxiv.org/abs/0904.4149}{{\ttfamily 0904.4149}}].

\bibitem{kra15}
J.~{Krause}, G.~{Morlino} and S.~{Gabici}, \emph{{CRIME - cosmic ray interactions in molecular environments}},  in \emph{34th International Cosmic Ray Conference (ICRC2015)}, vol.~34 of \emph{International Cosmic Ray Conference}, p.~518, July, 2015, \href{https://doi.org/10.22323/1.236.0518}{DOI}.

\bibitem{rav25}
S.~{Ravikularaman}, S.~{Recchia}, V.H.M.~{Phan} and S.~{Gabici}, \emph{{Cosmic rays cannot explain the high ionisation rates in the Galactic centre}}, \href{https://doi.org/10.1051/0004-6361/202451155}{\emph{A\&A} {\bfseries 694} (2025) A114} [\href{https://arxiv.org/abs/2406.15260}{{\ttfamily 2406.15260}}].

\bibitem{ber90}
V.S.~{Berezinskii}, S.V.~{Bulanov}, V.A.~{Dogiel} and V.S.~{Ptuskin}, \emph{{Astrophysics of cosmic rays}} (1990).

\bibitem{HESS2016}
{HESS Collaboration}, A.~{Abramowski}, F.~{Aharonian}, F.A.~{Benkhali}, A.G.~{Akhperjanian}, E.O.~{Ang{\"u}ner} et~al., \emph{{Acceleration of petaelectronvolt protons in the Galactic Centre}}, \href{https://doi.org/10.1038/nature17147}{\emph{Nature} {\bfseries 531} (2016) 476} [\href{https://arxiv.org/abs/1603.07730}{{\ttfamily 1603.07730}}].

\bibitem{HESS2006}
F.~{Aharonian}, A.G.~{Akhperjanian}, A.R.~{Bazer-Bachi}, M.~{Beilicke}, W.~{Benbow}, D.~{Berge} et~al., \emph{{Discovery of very-high-energy {\ensuremath{\gamma}}-rays from the Galactic Centre ridge}}, \href{https://doi.org/10.1038/nature04467}{\emph{Nature} {\bfseries 439} (2006) 695} [\href{https://arxiv.org/abs/astro-ph/0603021}{{\ttfamily astro-ph/0603021}}].

\bibitem{gag17}
D.~{Gaggero}, D.~{Grasso}, A.~{Marinelli}, M.~{Taoso} and A.~{Urbano}, \emph{{Diffuse Cosmic Rays Shining in the Galactic Center: A Novel Interpretation of H.E.S.S. and Fermi-LAT {\ensuremath{\gamma}} -Ray Data}}, \href{https://doi.org/10.1103/PhysRevLett.119.031101}{\emph{Phys. Rev. Lett.J} {\bfseries 119} (2017) 031101} [\href{https://arxiv.org/abs/1702.01124}{{\ttfamily 1702.01124}}].

\bibitem{law08}
C.J.~{Law}, F.~{Yusef-Zadeh}, W.D.~{Cotton} and R.J.~{Maddalena}, \emph{{Green Bank Telescope Multiwavelength Survey of the Galactic Center Region}}, \href{https://doi.org/10.1086/533587}{\emph{ApJS} {\bfseries 177} (2008) 255} [\href{https://arxiv.org/abs/0801.4294}{{\ttfamily 0801.4294}}].

\bibitem{str10}
A.W.~{Strong}, T.A.~{Porter}, S.W.~{Digel}, G.~{J{\'o}hannesson}, P.~{Martin}, I.V.~{Moskalenko} et~al., \emph{{Global Cosmic-ray-related Luminosity and Energy Budget of the Milky Way}}, \href{https://doi.org/10.1088/2041-8205/722/1/L58}{\emph{ApJL} {\bfseries 722} (2010) L58} [\href{https://arxiv.org/abs/1008.4330}{{\ttfamily 1008.4330}}].

\bibitem{cro11}
R.M.~{Crocker}, D.I.~{Jones}, F.~{Aharonian}, C.J.~{Law}, F.~{Melia}, T.~{Oka} et~al., \emph{{Wild at Heart: the particle astrophysics of the Galactic Centre}}, \href{https://doi.org/10.1111/j.1365-2966.2010.18170.x}{\emph{MNRAS} {\bfseries 413} (2011) 763} [\href{https://arxiv.org/abs/1011.0206}{{\ttfamily 1011.0206}}].

\bibitem{pon13}
G.~{Ponti}, M.R.~{Morris}, R.~{Terrier} and A.~{Goldwurm}, \emph{{Traces of Past Activity in the Galactic Centre}},  in \emph{Cosmic Rays in Star-Forming Environments}, D.F.~{Torres} and O.~{Reimer}, eds., vol.~34 of \emph{Astrophysics and Space Science Proceedings}, p.~331, Jan., 2013, \href{https://doi.org/10.1007/978-3-642-35410-6_26}{DOI} [\href{https://arxiv.org/abs/1210.3034}{{\ttfamily 1210.3034}}].

\bibitem{obo24}
M.~{Obolentseva}, A.V.~{Ivlev}, K.~{Silsbee}, D.A.~{Neufeld}, P.~{Caselli}, G.~{Edenhofer} et~al., \emph{{Reevaluation of the Cosmic-Ray Ionization Rate in Diffuse Clouds}}, \href{https://doi.org/10.3847/1538-4357/ad71ce}{\emph{ApJ} {\bfseries 973} (2024) 142} [\href{https://arxiv.org/abs/2408.11511}{{\ttfamily 2408.11511}}].

\end{thebibliography}\endgroup

\end{document}